\documentclass[sigconf]{acmart}

\usepackage{xcolor}
\usepackage{algpseudocode}
\usepackage{algorithmicx}
\usepackage{algorithm}
\usepackage{enumerate}
\usepackage{multirow}
\usepackage[shortlabels]{enumitem}
\usepackage{subcaption}

\usepackage{balance}

\def\BibTeX{{\rm B\kern-.05em{\sc i\kern-.025em b}\kern-.08emT\kern-.1667em\lower.7ex\hbox{E}\kern-.125emX}}
    
\copyrightyear{2019}
\acmYear{2019}
\setcopyright{acmcopyright}
\acmConference[CIKM '19] {The 28th ACM International Conference on Information and Knowledge Management}{November 3--7, 2019}{Beijing, China}
\acmBooktitle{The 28th ACM International Conference on Information and Knowledge Management (CIKM'19), November 3--7, 2019, Beijing, China}
\acmPrice{15.00}
\acmDOI{10.1145/3357384.3357879}
\acmISBN{978-1-4503-6976-3/19/11}

\MakeRobust{\Call}

\newcommand{\ouralgorithm}{FastRP}

\begin{document}

\fancyhead{}

%
\title{Fast and Accurate Network Embeddings via Very Sparse Random Projection}

%

\author{Haochen Chen}
\email{haocchen@cs.stonybrook.edu}
\affiliation{%
  \institution{Stony Brook University}
}

\author{Syed Fahad Sultan}
\email{ssyedfahad@cs.stonybrook.edu}
\affiliation{%
  \institution{Stony Brook University}
}

\author{Yingtao	Tian}
\email{yittian@cs.stonybrook.edu}
\affiliation{%
  \institution{Stony Brook University}
}

\author{Muhao Chen}
\email{muhaochen@ucla.edu}
\affiliation{%
  \institution{University of California, Los Angeles}
}

\author{Steven Skiena}
\email{skiena@cs.stonybrook.edu}
\affiliation{%
  \institution{Stony Brook University}
}

\begin{abstract}
We present \ouralgorithm{}, a scalable and performant algorithm for learning distributed node representations in a graph.
\ouralgorithm{} is over 4,000 times faster than state-of-the-art methods such as DeepWalk and node2vec,
while achieving comparable or even better performance as evaluated on several real-world networks on various downstream tasks.
We observe that most network embedding methods consist of two components:
construct a node similarity matrix and then apply dimension reduction techniques to this matrix.
We show that the success of these methods should be attributed to the
proper construction of this similarity matrix, rather than the dimension reduction method employed.

\ouralgorithm{} is proposed as a scalable algorithm for network embeddings.
Two key features of \ouralgorithm{} are: 1) it explicitly constructs a node similarity matrix
that captures transitive relationships in a graph and normalizes matrix entries based on node degrees;
2) it utilizes very sparse random projection, which is a scalable optimization-free method for dimension reduction.
An extra benefit from combining these two design choices is that it allows the iterative computation of node embeddings
so that the similarity matrix need not be explicitly constructed, which further speeds up \ouralgorithm.
\ouralgorithm{} is also advantageous for its ease of implementation, parallelization and hyperparameter tuning.
The source code is available at \url{https://github.com/GTmac/FastRP}.

\end{abstract}

%
%

\ccsdesc[500]{Information systems~Data mining}

%
\keywords{network embeddings; network representation learning; random projection}

%
\maketitle

\section{Introduction}

\begin{figure*}[t]
\centering
\begin{subfigure}[b]{.32\linewidth}
	\centering
    \includegraphics[width=\linewidth]{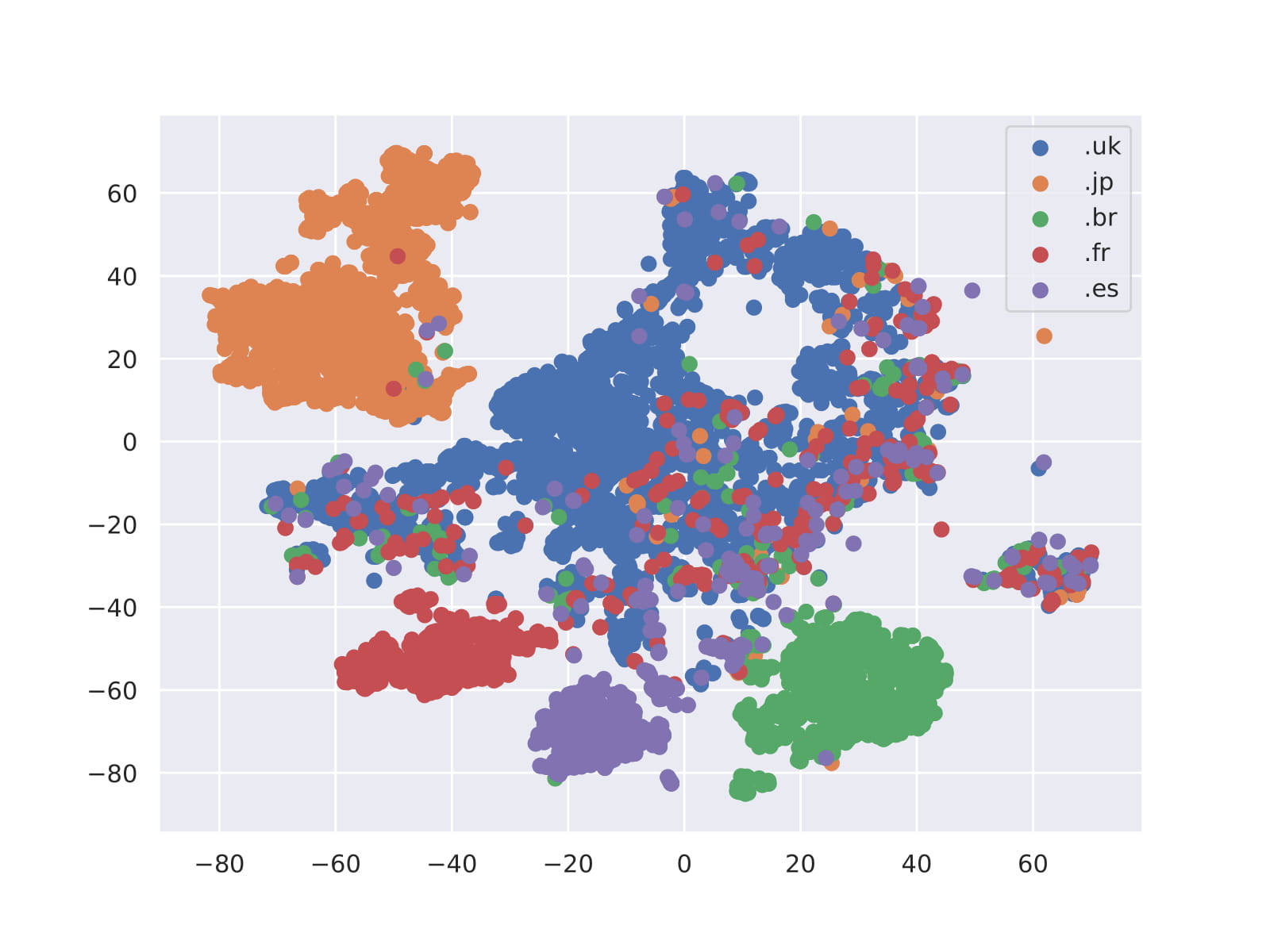}
    \caption{\ouralgorithm{}.}
    \label{fig:fastrp-tsne}
\end{subfigure}
\begin{subfigure}[b]{.32\linewidth}
	\centering
    \includegraphics[width=\linewidth]{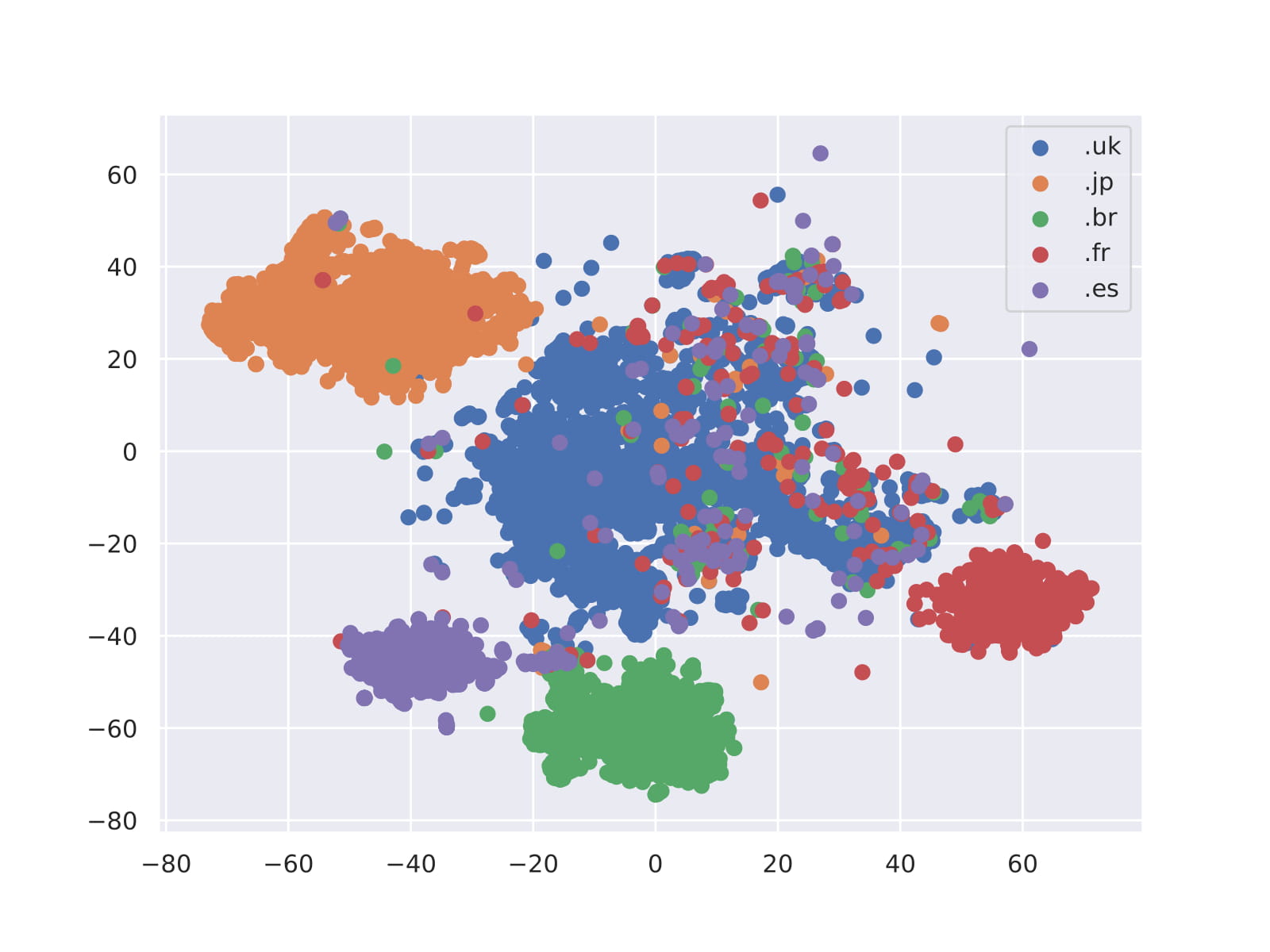}
    \caption{DeepWalk.}
    \label{fig:deepwalk-tsne}
\end{subfigure}
\begin{subfigure}[b]{.32\linewidth}
	\centering
    \includegraphics[width=\linewidth]{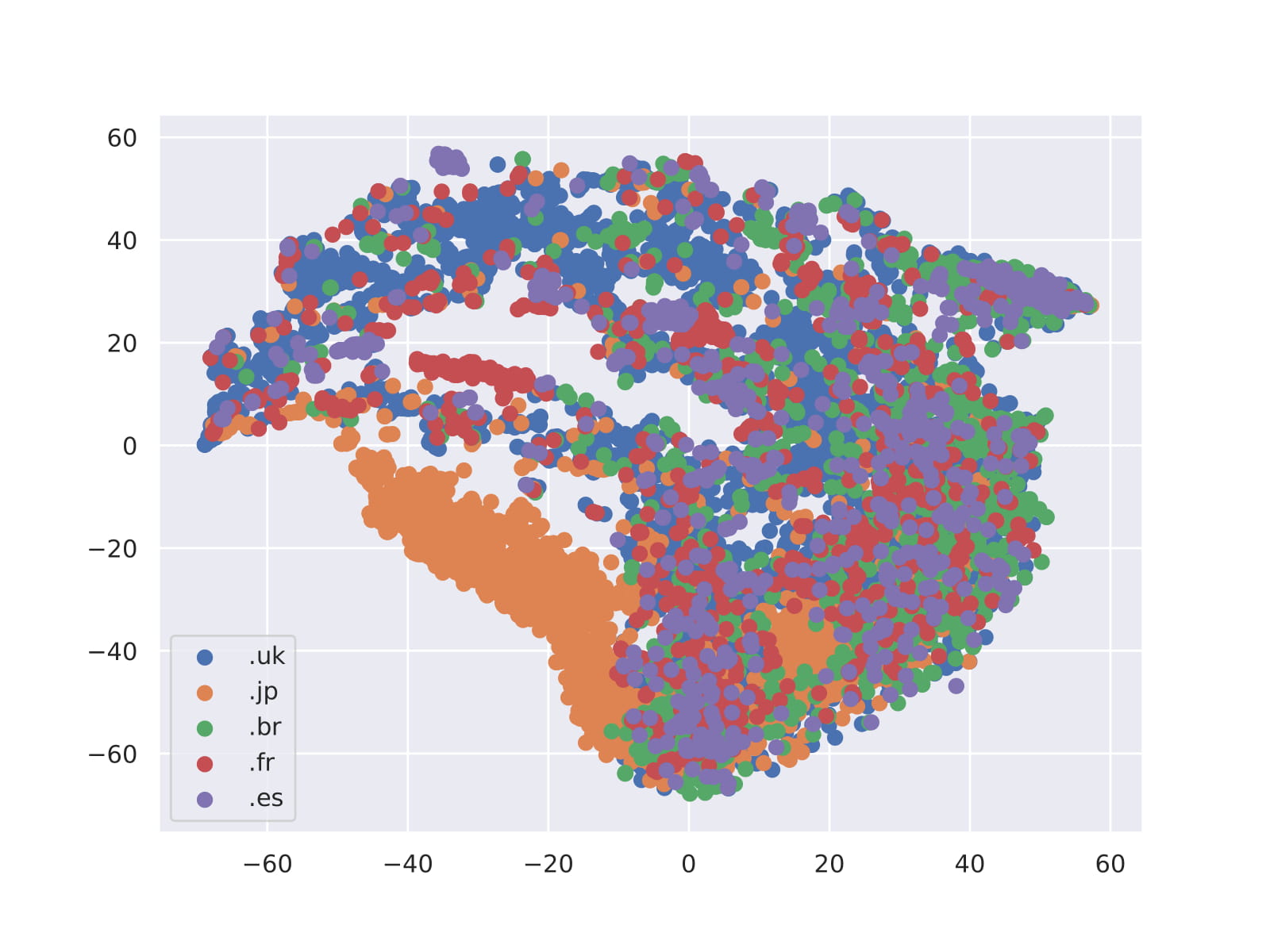}
    \caption{RandNE.}
    \label{fig:randne-tsne}
\end{subfigure}
\caption{Visualization of the embeddings produced by \ouralgorithm{}, DeepWalk and RandNE on the \texttt{WWW} network for websites from five country code top-level domains.
We use t-SNE to project the embeddings to two-dimensional space.}
\label{fig:vis}
\end{figure*}

Network embedding methods learn low-dimensional distributed representation of nodes in a network.
These learned representations can serve as latent features for a variety of inference tasks on graphs,
such as node classification~\cite{deepwalk}, link prediction~\cite{node2vec} and network reconstruction~\cite{sdne}.

Research on network embeddings dates back to early 2000s in the context of dimension reduction,
when methods such as LLE~\cite{lle}, IsoMap~\cite{isomap} and Laplacian Eigenmaps~\cite{le} were proposed.
These methods are general in that they embed an arbitrary $n \times m$ feature matrix ($n$ is the number of data points)
into an $n \times d$ embedding matrix, where $d \ll m$.
Although these methods produce high-quality embeddings, their time complexity is at least $O(n^2)$
which is prohibitive for large $n$.

More recent work in this area shifts their focus to embedding graph data,
which represents a special class of sparse feature matrix, where $n = m$.
The sparsity and discreteness of real-world graphs permit the design of more scalable network embedding algorithms.
The pioneering work here is DeepWalk~\cite{deepwalk}, which essentially samples node pairs from $k$-step transition matrices
with different values of $k$,
and then train a Skip-gram~\cite{skipgram} model on these pairs to obtain node embeddings.

The most significant contribution of DeepWalk is that it introduces a two-component paradigm for representation learning on graphs: first explicitly constructing a node similarity matrix or implicitly sampling node pairs from it, then performing dimension reduction on the matrix to produce node embeddings.
Much subsequent work has since followed to propose different strategies for both steps~\cite{line,node2vec,sdne,verse}.

Although most such methods are considered scalable with time complexity being linear to
the number of nodes and/or edges, we note that the constant factor is often too large to be ignored.
The reason is two-fold.
First, many of these methods are sampling-based and a huge number of samples is required to learn high-quality embeddings.
For example, DeepWalk samples about 32,000 context nodes per node under its default setting\footnote{We consider the recommended hyperparameter settings in the DeepWalk paper, where 80 random walks of length 40 are sampled per node and the window size for Skip-gram is 10.}.
Second, the dimension reduction methods being used also incur substantial computational cost,
making this constant factor even larger.
Popular methods such as DeepWalk~\cite{deepwalk}, LINE~\cite{line} and node2vec~\cite{node2vec} all adopt Skip-gram for learning node embeddings. However, optimizing the Skip-gram model is time-consuming due to the large number of gradient updates needed before the model converges.
As such, despite being the most scalable state-of-the-art network embedding algorithms, it still takes DeepWalk and LINE several days of CPU time to embed the Youtube graph~\cite{tang2009scalable}, a moderately sized graph with 1M nodes and 3M edges.

Can we design a truly scalable network embedding method that produces node embeddings for million-scale graphs in several minutes without compromising the representation quality?
To answer this question, we analyze several state-of-the-art network embedding methods by examining their design choices for both similarity matrix construction and dimension reduction.
Our analysis motivates us to propose \ouralgorithm{}, which presents much more scalable solutions to both steps without compromising embedding quality.

To illustrate the effectiveness of \ouralgorithm{}, we visualize the node representations produced by \ouralgorithm{}, DeepWalk~\cite{deepwalk} and an earlier random projection-based method, RandNE~\cite{randne} on the WWW network (Figure~\ref{fig:vis}).
The nodes are hostnames such as \texttt{youtube.com} and \texttt{instagram.com}.
For the purpose of visualization, we use t-SNE~\cite{tsne} to project the node embeddings to two-dimensional space.
We take the websites from five countries: United Kingdom (\texttt{.uk}), Japan (\texttt{.jp}), Brazil (\texttt{.br}), France (\texttt{.fr}) and Spain (\texttt{.es}) as indicated by the color of the dots.
Observe that for \ouralgorithm{} and DeepWalk, the websites from each top-level domain form clusters that are very well separated.
For the RandNE embeddings, there is no clear boundary between the websites from different top-level domains.
\ouralgorithm{} achieves similar quality to DeepWalk while being over 4,000 times faster.

To sum up, our contributions are the following:

	\noindent \textbf{Improved Understanding of Existing Network Embedding Algorithms.}
	By viewing representative network embedding algorithms as a procedure with two components, similarity matrix construction and dimension reduction, we gain an improved understanding of why these algorithms work and why do they have scalability issues.
	This improved understanding motivates us to propose new solutions for both components.
	
	\noindent \textbf{Better Formulation of the Node Similarity Matrix.}
	We construct a node similarity matrix with two unique properties:
	1) it considers the implicit, transitive relationships between nodes;
	2) it normalizes pairwise similarity of nodes based on node degrees.
	
	\noindent \textbf{More Scalable Dimension Reduction Algorithm.}
	Different from previous work that relies on time-consuming dimension reduction methods such as Skip-gram and SVD, we obtain node embeddings via very sparse random projection of the node similarity matrix.
	An additional benefit from combining these two design choices is that it allows the iterative computation of node embeddings, which has linear cost in the size of the graph.
	
	\noindent \textbf{DeepWalk Quality Embeddings that is Produced Over 4,000 Times Faster}. Extensive experimental results show that \ouralgorithm{} produces high-quality
	node embeddings comparable to state-of-the-art methods while being at least three orders of magnitude faster.

\section{Preliminaries}

In this section, we give the formal definition of network embeddings and
introduce the paradigm of network embeddings as a two-component process.
We then detail the design decisions of several state-of-the-art methods for both components and show why they have scalability issues.

\subsection{Notation and Task Definition}
We consider the problem of embedding a network: given an undirected graph,
the goal is to learn a low-dimensional latent representation for each node in the graph\footnote{We use network and graph interchangeably.}.
Formally, let $G = (V, E)$ be a graph, where $V$ is the set of nodes and $E$ is the set of edges.
Let $n = |V|$ be the number of nodes, $m = |E|$ be the number of edges, $d_i$ be the degree of the $i$-th node, and $\mathbf{S}$ be the adjacency matrix of $G$.
The goal of network embeddings is to develop a mapping $\Phi: V \mapsto \mathbf{N} \in \mathbb{R}^{n \times d}, d \ll n$.
For a node $v \in V$, we call the $d$-dimensional vector $\mathbf{N}_{v}$ its embedding vector (or node embedding).

Network embeddings can be viewed as performing \textit{dimension reduction} on graphs:
the input is an $n \times n$ feature matrix associated with the graph,
on which we apply dimension reduction techniques to reduce its dimensionality to $n \times d$.
This leads to two questions:
\begin{enumerate}
    \item What is an appropriate node similarity matrix to perform dimension reduction on?
    \item What dimension reduction techniques should be used?
\end{enumerate}

We now review the existing solutions to both questions.

\subsection{Node Similarity Matrix}
The most straightforward input matrices to consider is the adjacency matrix $\mathbf{S}$ or the transition matrix $A$:
\[
\mathbf{A} = \mathbf{D}^{-1}\mathbf{S}
\]
where $\mathbf{D}$ is the degree matrix of $G$:
\[
    \mathbf{D}_{ij}=\left\{\begin{array}{ll}{\sum_{p} \mathbf{S}_{ip}} & {\text { if } i=j,} \\ {0} & {\text {otherwise.}}\end{array}\right.
\]
However, directly applying dimension reduction techniques on $S$ or $A$ is problematic.
Real-world graphs are usually extremely sparse, which means most of the entries in $S$ are zero.
However, the absence of edge between two nodes $u$ and $v$ does not imply that there is no association between them.
In particular, if two nodes are not adjacent but are connected by a large number of paths,
it is still likely that there is a strong association between them.

The intuition above motivates us to exploit higher-order relationships in the graph.
A natural high-order node similarity matrix is the $k$-step transition matrix:
\begin{equation}
    \label{eq:a-k}
    \mathbf{A}^{k}=\underbrace{\mathbf{A} \cdots \mathbf{A}}_{k}
\end{equation}
The $ij$-th entry of $\mathbf{A}^k$ denotes the probability of reaching $j$ from $i$ in exactly $k$ steps of random walk.
We will show that many existing methods adopt variations of this definition of similarity matrix.

\subsection{Dimension Reduction Techniques}
\label{sec:dim-reduction}
Once an $n \times n$ similarity matrix is constructed, network embedding methods perform dimension reduction on it to obtain node representations.
In this section, we introduce two commonly used dimension reduction techniques: singular value decomposition (SVD) and Skip-gram.

\noindent \textbf{SVD}. SVD~\cite{truncated-svd} is a classical matrix factorization method for dimension reduction. SVD factorizes a feature matrix $M$ into the product of three matrices: $\mathbf{M} = \mathbf{U} \cdot \mathbf{\Sigma} \cdot \mathbf{V}^{\top}$,
where $U$ and $V$ are orthonormal and $\Sigma$ is a diagonal matrix consisting of singular values.
To perform dimension reduction with SVD, it is common to take the top $d$ singular values $\Sigma_d$ from $\Sigma$ and the corresponding columns from $U$ and $V$.

\noindent \textbf{Skip-gram}.
Skip-gram~\cite{skipgram} is a method for learning word embeddings,
which is also shown to be performant in the context of network embeddings.
Skip-gram works by sampling word pairs (or node pairs) from a word co-occurrence matrix (or node similarity matrix) $C$ and modeling the probability that a given word-context pair $(u, v)$ is from $C$ or not.
Goldberg and Levy~\cite{word-emb-as-mf} showed that Skip-gram is implicitly factorizing a shifted pointwise mutual information (PMI) matrix of word co-occurrences.
Formally, the matrix Skip-gram seeks to factorize has elements:
\begin{equation}
\label{eq:sgns-mf}
\mathbf{M}_{uv} =\log \frac{\#(u, v) \cdot|C|}{\#(u) \cdot \#(v)} - \log b = \log \left(\mathrm{PMI} (u, v) \right) - \log b
\end{equation}
where $\#(u)$ denotes the occurrence count of $u$ in $C$ and $b$ denotes the number of negative samples in Skip-gram.

\subsection{Representative Network Embedding Methods}
In this section, we discuss how three representative network embedding methods: DeepWalk~\cite{deepwalk}, LINE~\cite{line} and GraRep~\cite{grarep} fit into the two-component procedure described above. The analysis of node2vec~\cite{node2vec} is similar to that of DeepWalk, which we omit here.

\noindent \textbf{DeepWalk}~\cite{deepwalk}.
DeepWalk's core idea is to sample node pairs from a weighted combination of $A, A^2, \cdots, A^k$, and then train a Skip-gram model on these samples.
Making use of Eq.~\ref{eq:sgns-mf}, it can be shown that DeepWalk is implicitly factorizing the following matrix~\cite{netmf}:
\begin{equation}
    \label{eq:dw-mf}
    \mathbf{M} = \log \left(\operatorname{vol}(G) \left(\frac{1}{k} \sum_{r=1}^{k}\mathbf{A}^{r} \mathbf{D}^{-1} \right) \right) - \log b
\end{equation}
where $\operatorname{vol}(G) = \sum_{i}\sum_{j} S_{ij}$.

\noindent \textbf{LINE}~\cite{line}.
LINE can be seen as a variation of DeepWalk that only considers node pairs
that are at most two hops away.
Using a similar derivation, it can be shown that LINE implicitly factorizes:
\begin{equation}
    \mathbf{M} = \log \left( \operatorname{vol}(G) \left( \mathbf{A} \mathbf{D}^{-1} \right) \right) - \log b
\end{equation}

\noindent \textbf{GraRep}~\cite{grarep}.
GraRep can be regarded as the matrix factorization version of DeepWalk.
Instead of sampling from $A, A^2, \cdots, A^k$, it directly computes these matrices and then factorizes the corresponding shifted PMI matrix for each power of $A$.

\subsection{Scalability of Representative Methods}

Putting existing methods into this two-component framework reveals their intrinsic scalability issues as following: 

\noindent \textbf{Node Similarity Matrix Construction.}
Many previous studies have demonstrated the importance of preserving high-order proximity between nodes~\cite{deepwalk,grarep,netmf,arope,randne}, which is typically done by raising $\mathbf{A}$ to $k$-th power and optionally normalize it afterward (see Eq.~\ref{eq:dw-mf} for an example).
This causes scalability issues since both computing $\mathbf{A}^k$ and applying a transformation to each element in $\mathbf{A}^k$ are at least quadratic.
For methods such as DeepWalk and node2vec, this problem is slightly mitigated by sampling node pairs from $\mathbf{A}^k$ instead.
But still, a huge number of samples is required for them to get an accurate enough estimation of $\mathbf{A}^k$.

\noindent \textbf{Dimension Reduction.} The dimension reduction techniques adopt\-ed by these methods also affect their scalability. Both Skip-gram and SVD are not among the fastest dimension reduction algorithms~\cite{random-projection}.

In the next section, we present our solutions to both problems that allow for better scalability.

\section{Method}

In this section, we introduce \ouralgorithm{}.
We first describe the usage of very sparse random projection for dimension reduction and its merit in preserving high-order proximity.
Then, we present our design of the node similarity matrix.
This matrix is carefully designed so that: 1) it preserves transitive relationships in the input graph; 2) its entries are properly normalized; 3) it can be formulated as matrix chain multiplication, so that applying random projection on this matrix only costs linear time.
Lastly, we discuss several additional advantages of \ouralgorithm{}.

\subsection{Very Sparse Random Projection}
Random projection is a dimension reduction method that preserves pairwise distances between data points with strong theoretical guarantees~\cite{random-projection}.
The idea behind this is very simple: to reduce an $n \times m$ (for graph data, we have $n = m$) feature matrix $\mathbf{M}$ to an $n \times d$ matrix $\mathbf{N}$ where $d \ll m$, we can simply multiply the feature matrix with an $m \times d$ random projection matrix $\mathbf{R}$:
\begin{equation}
    \label{eq:rp}
    \mathbf{N} = \mathbf{M} \cdot \mathbf{R}
\end{equation}
As long as the entries of $\mathbf{R}$ are i.i.d with zero mean, $\mathbf{N}$ is able to preserve the pairwise distances in $\mathbf{A}\mathbf{A^{\top}}$~\cite{arriaga1999algorithmic}.

The difference among different random projection algorithms is mostly in the construction of $\mathbf{R}$.
The most studied one is Gaussian random projection, where entries of $\mathbf{R}$ are sampled i.i.d. from a Gaussian distribution: $\mathbf{R}_{ij} \sim \mathcal{N}\left(0, {1}/{d}\right)$.
Since $\mathbf{R}$ is a dense $m \times d$ matrix, the time complexity of Gaussian random projection is $O(n \cdot m \cdot d)$.

As an improvement to Gaussian random projection, Achlioptas~\cite{sparse-rp} proposed \textit{sparse random projection}, where entries of $\mathbf{R}$ are sampled i.i.d. from
\begin{equation}
    \label{eq:sparse-rp}
    \mathbf{R}_{ij} = \left\{\begin{aligned} \sqrt{s} & \text { with probability } \frac{1}{2 s} \\ 0 & \text { with probability } 1-\frac{1}{s} \\ -\sqrt{s} & \text { with probability } \frac{1}{2 s} \end{aligned}\right.
\end{equation}
where $s=3$ is used.
This leads to a 3x speedup since ${2}/{3}$ of the entries of $\mathbf{R}$ are zero.
Additionally, this configuration of $\mathbf{R}$ does not require any floating-point computation since the multiplication with $\sqrt{s}$ can be delayed, providing additional speedup.

Li et al.~\cite{very-sparse-rp} extend Achlioptas~\cite{sparse-rp} by showing that $s \gg 3$ can be used to further speed up the computation.
They recommend setting $s=\sqrt{m}$, which achieves $\sqrt{m}$ times speedup over Gaussian random projection while ensuring the quality of the embeddings.
In this work, we consider this \textit{very sparse random projection} method for dimension reduction of the node similarity matrix.

As an optimization-free dimension reduction method, very sparse random projection wins over SVD and Skip-gram for its superior computational efficiency.
The fact that it only requires matrix multiplication also enables faster computation on accelerators such as GPUs, as well as easy parallelization.

Apart from these advantages, the random projection approach also benefits from the \textit{associative} property of matrix multiplication.
To see why this is important, consider the basic form of high-order similarity matrix $\mathbf{A}^k$, as defined in Eq.~\ref{eq:a-k}.
To compute its random projection $\mathbf{N} = \mathbf{A}^k \cdot \mathbf{R}$, there is no need to calculate $\mathbf{A}^k$ from scratch since the computation can be done iteratively:
\begin{equation}
    \label{eq:iterative-rp}
    \mathbf{N} = (\underbrace{\mathbf{A} \cdot \cdots (\mathbf{A} \cdot (\mathbf{A}}_{k} \cdot \mathbf{R}) ) )
\end{equation}
This reduces the time complexity from $O(n^3 \cdot k \cdot d)$ to $O(m \cdot k \cdot d)$.

\subsection{Similarity Matrix Construction}
The next step is to construct a proper node similarity matrix leveraging the associative property of matrix multiplication.
We make two key observation about the similarity matrices used by the existing method.
First, it is important to preserve high-order proximity in the input graph -- this is typically done by raising $\mathbf{A}$ to $k$-th power.
Second, element-wise normalization is performed (taking logarithm for DeepWalk, LINE and GraRep) on the raw similarity matrix before dimension reduction.

Most previous matrix-based network embedding methods emphasize on the importance of high-order proximity but skip the element-wise normalization step for either better scalability or ease of analysis~\cite{netmf,netsmf,randne}.
Is normalization of the node similarity matrix important? If so, is there any other normalization method that allows for scalable computation?
We answer these questions by analyzing the properties of $\mathbf{A}^k$ from a spectral graph theory perspective.

To begin with, we consider a transformation of $\mathbf{A}$ defined as
$\mathbf{B} = \mathbf{D}^{\frac{1}{2}}\mathbf{A}\mathbf{D}^{-\frac{1}{2}}$.
Since $\mathbf{B}$ is a real symmetric matrix, it can be decomposed as
$\mathbf{B}=\mathbf{Q} \mathbf{\Lambda} \mathbf{Q}^{\top}$
where $\mathbf{\Lambda}$ is a diagonal matrix of eigenvalues $\lambda_1 \geq \lambda_2 \geq \ldots \geq \lambda_n$ of $\mathbf{B}$, and $\mathbf{Q}$ is an orthogonal matrix consisting of the corresponding eigenvectors $\mathbf{q}_1, \ldots, \mathbf{q}_n$.

It is easy to verify that $\left(1, \mathbf{w} = \left( \sqrt{d_1}, \cdots, \sqrt{d_n} \right) \right)$ is an eigenpair of $\mathbf{B}$.
Following the Frobenius-Perron Theorem~\cite{fp-theorem}, we have:
\[
    \lambda_{1}=1 > \lambda_{2} \geq \ldots \geq \lambda_{n} \geq -1
\]
and
\[
\mathbf{q}_1 = \left( \sqrt{\frac{d_1}{2m} }, \cdots, \sqrt{\frac{d_n}{2m} } \right)
\]
Now:
\[
\begin{aligned}
    \mathbf{A}^{k} & = \mathbf{D}^{-\frac{1}{2}} \mathbf{B}^{k} \mathbf{D}^{\frac{1}{2}}
    = \mathbf{D}^{-\frac{1}{2}} (\mathbf{Q}\mathbf{\Lambda}\mathbf{Q}^{\top})^{k} \mathbf{D}^{\frac{1}{2}} \\
    & = \mathbf{D}^{-\frac{1}{2}} \mathbf{Q} \mathbf{\Lambda}^{k} \mathbf{Q}^{\top} \mathbf{D}^{\frac{1}{2}} \\
    & = \sum_{t=1}^{n}\lambda_{t}^k \mathbf{D}^{-\frac{1}{2}} \mathbf{q}_t \mathbf{q}_t^{\top} \mathbf{D}^{\frac{1}{2} }\\
    & = \mathbf{P} + \sum_{t=2}^{n}\lambda_{t}^k \mathbf{D}^{-\frac{1}{2}} \mathbf{q}_t \mathbf{q}_t^{\top} \mathbf{D}^{\frac{1}{2} }
\end{aligned}
\]
where $\mathbf{P}_{ij} = {d_j}/{2m}$.

For a particular entry $\mathbf{A}_{ij}^{k}$ we have:
\begin{equation}
    \mathbf{A}_{ij}^{k} = \frac{d_j}{2m} +
    \sum_{t=2}^{n}\lambda_t^k q_{ti}q_{tj} \sqrt{\frac{d_j}{d_i}}
\end{equation}

This derivation illustrates the importance of normalization.
Since $|\lambda_t| < 1$ holds for $t = 2, \ldots, n$ (assuming $G$ is non-bipartite), we have $\mathbf{A}_{ij}^{k} \rightarrow {d_j}/{2m}$ when $k \rightarrow \infty$.
Since many of the real-world graphs are scale-free~\cite{scale-free}, it follows that the entries in $\mathbf{A}^k$ also has a heavy-tailed distribution.

The heavy-tailed distribution of data causes problems for dimension reduction methods~\cite{very-sparse-rp}.
The pairwise distances between data points are dominated by the columns with exceptionally large values, rending them less meaningful.
In practice, term weighting schemes are applied to heavy-tailed data to reduce its kurtosis and skewness~\cite{salton1988term}.
Here, we consider a scaled version of the Tukey transformation~\cite{tukey}.
Concretely, we transform a feature $y$ into $y^\lambda$, where $\lambda$ controls the strength of normalization.
Now the only problem is that the exact feature values in $\mathbf{A}^k$ are not known, and we do not want to calculate these values for better scalability.
But again, we can rely on the fact that $\mathbf{A}_{ij}^{k}$ converges to ${d_j}/{2m}$.
The normalization we consider is therefore:
\begin{equation}
    \label{eq:normalization}
    \tilde{\mathbf{A}}_{ij}^{k} = \mathbf{A}_{ij}^{k} \cdot \left(\frac{d_j}{2m}\right)^{\lambda - 1}
    \approx {\mathbf{A}}_{ij}^{k} \cdot \left(\mathbf{A}_{ij}^{k}\right)^{\lambda - 1} \approx \left(\mathbf{A}_{ij}^{k}\right)^{\lambda}
\end{equation}

\subsection{Our Algorithm: \ouralgorithm{}}

\begin{algorithm}[t]
\begin{algorithmic}[1]
\Require
\Statex graph transition matrix $\mathbf{A}$, embedding dimensionality $d$, maximum power $k$,
normalization strength $\beta$, weights $\alpha_1, \alpha_2, \ldots, \alpha_k$
\Ensure matrix of node representations $\mathbf{N} \in \mathbb{R}^{n \times d}$
    \State Produce $\mathbf{R} \in \mathbb{R}^{n \times d}$ according to Eq.~\ref{eq:sparse-rp}
    \State $\mathbf{N}_1 \leftarrow \mathbf{A} \cdot \mathbf{L} \cdot \mathbf{R}$ where $\mathbf{L}_{ij} = \left( \frac{d_j}{2m} \right)^{\beta}$
    \For{$i=2$ to $n$}
        \State $\mathbf{N}_i \leftarrow \mathbf{A} \cdot \mathbf{N}_{i - 1}$
    \EndFor
	\State $\mathbf{N} = \alpha_1 \mathbf{N_1} + \ldots + \alpha_{k}\mathbf{N}_k$
    \State \Return $\mathbf{N}$
\end{algorithmic}
\caption{\ouralgorithm($\mathbf{A}$)}
\label{alg:fastrp}
\end{algorithm}

Let $\beta=\lambda - 1$, the normalization scheme in Eq.~\ref{eq:normalization} can be represented in matrix form:
$\tilde{\mathbf{A}}^{k} = \mathbf{A}^{k} \cdot \mathbf{L}$
where $\mathbf{L} = \mathrm{diag} \left( \left(\frac{d_1}{2m}\right)^{\beta}, \ldots, \left(\frac{d_n}{2m}\right)^{\beta} \right)$.
This allows for matrix chain multiplication when performing random projection:
$$
    \mathbf{N} = \tilde{\mathbf{A}}^{k} \cdot \mathbf{R} = (\underbrace{\mathbf{A} \cdot \cdots (\mathbf{A} \cdot (\mathbf{A}}_{k} \cdot \mathbf{L} \cdot \mathbf{R}) ) )
$$
We further consider a weighted combination of different powers of $\mathbf{A}$, so that the embeddings of $G$ is computed as follows:
$$
    \mathbf{N} = \left(\alpha_1 \tilde{\mathbf{A}} + \alpha_2 \tilde{\mathbf{A}}^2 + \ldots +
    \alpha_k \tilde{\mathbf{A}}^k \right) \cdot \mathbf{R}
$$
where $\alpha_1, \alpha_2, \ldots, \alpha_k$ are the weights.
The outline of \ouralgorithm{} is presented in Algorithm~\ref{alg:fastrp}.

\subsection{Time Complexity}
\label{sec:time-complexity}
The time complexity of \ouralgorithm{} is $O\left({(n \cdot d)}/{s}\right) = O\left(n \cdot \sqrt{d}\right)$ for constructing the sparse random projection matrix (line 1),
$O(m \cdot k \cdot d)$ for random projection (line 2 to 5) for each power of $\mathbf{A}$
and $O(n \cdot k \cdot d)$ for merging embedding matrices (line 6).
Overall, the time complexity of \ouralgorithm{} is $O( (n + m) \cdot k \cdot d)$, which is linear to the number of nodes and edges in $G$.

\subsection{Implementation, Parallelization and Hyperparameter Tuning}
\label{sec:tuning}
\noindent \textbf{Implementation.}
The implementation of \ouralgorithm{} is very simple and straightforward, with less than 100 lines of Python code.

\noindent \textbf{Parallelization.}
Besides the ease of implementation, our algorithm is also easy to parallelize, since the only operation involved is matrix multiplication.
One easy way to speed up matrix multiplication $\mathbf{C} = \mathbf{A} \cdot \mathbf{B}$ is to perform block partitioning on the input matrices $\mathbf{A}$ and $\mathbf{B}$:
\begin{equation}
    A=\left( \begin{array}{ll}{A_{11}} & {A_{12}} \\ {A_{21}} & {A_{22}}\end{array}\right), \quad
    B=\left( \begin{array}{ll}{B_{11}} & {B_{12}} \\ {B_{21}} & {B_{22}}\end{array}\right)
\end{equation}
Then it is easy to see that:
\begin{equation}
    C = \left( \begin{array}{cc}{A_{11} B_{11}+A_{12} B_{21}} & {A_{11} B_{12}+A_{12} B_{22}} \\ {A_{21} B_{11}+A_{22} B_{21}} & {A_{21} B_{12}+A_{22} B_{22}}\end{array}\right)
\end{equation}
The recursive matrix multiplications and summations can be performed in parallel.
The smaller block matrices $A_{11}, A_{12}, \cdots, B_{11}, B_{12}, \cdots$ can be further partitioned recursively for execution on more processors~\cite{cilk}.

\noindent \textbf{Efficient Hyperparameter Tuning.}
\ouralgorithm{} also allows for highly efficient hyperparameter tuning.
The idea is to first pre-compute the embeddings $\mathbf{N}_1, \mathbf{N}_2, \cdots, \mathbf{N}_k$ derived from different orders of proximity matrices.
Since the final embedding matrix is a weighted combination of $\mathbf{N}_1, \mathbf{N}_2, \cdots, \mathbf{N}_k$, we only need to perform weighted summation during hyperparameter optimization.
Furthermore, according to the Johnson-Lindenstrauss lemma~\cite{jl-proof}, the embedding dimensionality $d$ determines the approximation error of random projection.
The implication of this is that we can efficiently tune hyperparameters on smaller values of $d$.
This is in contrast to most of the existing algorithms, which require retraining of the entire model for each hyperparameter configuration.

Besides these merits, the biggest advantage of \ouralgorithm{} is its superior computational efficiency.
In the next section, we will show that \ouralgorithm{} is orders of magnitude faster than the state-of-the-art methods while achieving comparable or even better performance.

\section{Experiments}
In this section, we conduct experiments to evaluate the performance of \ouralgorithm{}.
We first provide an overview of the datasets.
Then, we compare \ouralgorithm{} with a number of baseline methods both in terms of running time and performance on downstream tasks.
We further discuss the performance of our method with regard to several important hyperparameters and its scalability.

\subsection{Datasets}

\begin{table*}
	\centering
	{
    \begin{tabular}{l@{\quad}c c c c c}
    \toprule
    Name & \# Vertices & \# Edges & \# Classes & Task\\
    \midrule
    WWW-200K & 200,000 & 32,822,166 & - & K-Nearest Neighbors \\
    WWW-10K & 10,000 & 3,904,610 & 50 & Node Classification\\
    Blogcatalog & 10,312 & 333,983 & 39 & Node Classification\\
    Flickr & 80,513 & 5,899,882 & 195 & Node Classification\\
    \bottomrule
    \end{tabular}
    }
    \caption{Statistics of the graphs used in our experiments.}
    \label{tab:dataset-stats}
\end{table*}
Table~\ref{tab:dataset-stats} gives an overview of the datasets used in experiments.
\begin{itemize}
    \item WWW-200K and WWW-10K~\cite{www-data}: these graphs are derived from the Web graph provided by Common Crawl, where the nodes are hostnames and the edges are the hyperlinks between these websites.
    For simplicity, we treat this graph as an undirected graph.
    The original graph has 385 million nodes and 2.5 billion edges, which is too large to be loaded into the memory of our machine. Thus, we construct subgraphs of this graph by taking the top 200,000 and 10,000 websites respectively as ranked by Harmonic Centrality~\cite{centrality}.
    We also use the \texttt{WWW-10K} graph for node classification, for which the label of a node is its top-level domain name such as \texttt{.org}, \texttt{.edu} and \texttt{.es}.
    \item Blogcatalog~\cite{tang2009scalable}: this is a network between bloggers on the Blogcatalog website.
    The labels are the categories a blogger publishes in.
    \item Flickr~\cite{tang2009scalable}: this is a network between the users on the photo sharing website Flickr.
    The labels represent the interest groups a user joins.
\end{itemize}

\subsection{Baseline Methods}
We compare \ouralgorithm{} against the following baseline methods:
\begin{itemize}
    \item DeepWalk~\cite{deepwalk} -- DeepWalk is a network embedding method that samples short random walks from the input graph.
    Then, these random walks are fed into the Skip-gram model to produce node embeddings.
    \item node2vec~\cite{node2vec} -- node2vec extends DeepWalk by performing biased random walks that balance between DFS and BFS.
    It also adopts SGNS as the dimension reduction method.
    \item LINE~\cite{line} -- LINE samples node pairs that are either adjacent or two hops away.
    LINE adopts Skip-gram with negative sampling (SGNS) to learn network embeddings from the samples.
    \item RandNE~\cite{randne} -- RandNE constructs a node similarity matrix that preserves the high-order proximity by raising the adjacency (or transition) matrix to the $k$-th power.
    Then, node embeddings are obtained by applying Gaussian random projection to this matrix.
\end{itemize}
We realize that there are many other recently proposed network embedding methods.
We do not include these methods as baselines since their performance are generally inferior to DeepWalk according to a recent comparative study~\cite{ne-comparison}.
Moreover, many of them are not scalable~\cite{ne-comparison}.

\subsection{Parameter Settings}
Here we present the parameter settings for the baseline models and our model.

\textbf{\ouralgorithm{} }. For \ouralgorithm{}, we set embedding dimensionality $d$ to 512 and maximum power $k$ to 4.
For the weights $\alpha_1, \alpha_2, \cdots, \alpha_k$, we observe that simply use a weighted combination of $\mathbf{A}^3$ and $\mathbf{A}^4$ is already enough for achieving competitive results.
Thus, we set $\alpha_1, \alpha_2, \alpha_3$ to $0, 0, 1$ respectively and tune $\alpha_4$.

Overall, we only have two hyperparameters to tune: normalization strength $\beta$ and the weight $\alpha_4$ for $\mathbf{A}^4$.
We use optuna\footnote{https://github.com/pfnet/optuna}, a Bayesian hyperparameter optimization framework to tune them.
The hyperparameter optimization is performed for 20 rounds on a small validation set of $1\%$ labeled data; the search ranges for $\beta$ and $\alpha_4$ are set to $\left[-1, 0\right]$ and $[2^{-3}, 2^6]$ respectively.
We also use a lower embedding dimensionality of $d=64$ to speed up the tuning process.

\textbf{RandNE}.
For RandNE, we set embedding dimensionality $d$ to 512 and maximum order $q$ to 3.
We note that for RandNE, incorporating the embeddings from $\mathbf{A}^4$ does not improve the quality of embeddings according to our experiments.
To ensure a fair comparison, we also conduct hyperparameter search for the weights in RandNE using the same procedure as \ouralgorithm{}.
The only difference is that instead of tuning $\beta$ and $\alpha_4$, we optimize the weights of $\mathbf{A}^2$ and $\mathbf{A}^3$.

\textbf{DeepWalk}.
For DeepWalk,  we need to set the following parameters: the number of random walks $\gamma$,
walk length $t$, window size $w$ for the Skip-gram model and representation size $d$.
We adopt the hyperparameter settings recommended in the original paper: $\gamma=80, t = 40, w = 10, d = 128$.

\textbf{node2vec}.
Since node2vec is built upon DeepWalk, we use the same parameter settings for node2vec as DeepWalk: $\gamma=80, t=40, w=10, d=128$.
We notice that these parameter settings lead to better results than the default settings as described in the paper, possibly because the total number of samples is larger.
For the in-out parameter $p$ and return parameter $q$, we conduct grid search over $p, q \in \{0.25, 0.50, 1, 2, 4\}$ as suggested in the paper.

\textbf{LINE}.
We use LINE with both the first order and second order proximity with the recommended hyperparameters.
Concretely, we set the dimensionality of embeddings to 200, the number of node pair samples to 10 billion and the number of negative samples to 5.

All the experiments are conducted on a single machine with 128\ GB memory and 40 CPU cores at 2.2\ GHz.
We note that \ouralgorithm{} and all the baseline methods support multi-threading.
However, for a fair running time comparison, we run all methods with a single thread and measure the CPU time (process time) consumed by each method.

\subsection{Runtime Comparison}
We first showcase the superior efficiency of our method by reporting the CPU time of \ouralgorithm{} and the baseline methods on all datasets in Table~\ref{tab:runtime}.
\ouralgorithm{} achieves at least 4,000x speedup over the state-of-the-art method DeepWalk.
For example, it takes \ouralgorithm{} less than 3 minutes to embed the \texttt{WWW-200K} graph, whereas DeepWalk takes almost a week to finish.
Node2vec is even slower; although LINE is several times faster than DeepWalk and node2vec, it is still a few hundreds of times slower than \ouralgorithm{}.
The only method with comparable running time is RandNE which uses Gaussian random projection for dimension reduction, but it is also slightly slower than \ouralgorithm{}.
Moreover, in the experiments below, we will show that the quality of embeddings produced by \ouralgorithm{} is significantly better than that of RandNE.

\begin{table*}[t!]
    \centering
    \begin{tabular}{ c r r r r r r}
    \toprule
        \textbf{Dataset} & \multicolumn{5}{c}{\textbf{Algorithm}} & Speedup over\\
        & FastRP & RandNE & LINE & DeepWalk & node2vec & DeepWalk \\ \midrule
        \texttt{WWW-200K} & 136.0 seconds & 169.8 seconds & 4.6 hours & 6.9 days & 63.8 days & \textbf{4383x} \\
        \texttt{WWW-10K} & 7.8 seconds & 13.6 seconds & 3.2 hours & 9.2 hours & 59.8 hours & \textbf{4246x} \\
        \texttt{Blogcatalog} & 6.0 seconds & 10.5 seconds & 3.0 hours & 8.7 hours & 41.2 hours & \textbf{5220x} \\
        \texttt{Flickr} & 33.1 seconds & 45.1 seconds & 4.2 hours & 3.1 days & 28.5 days & \textbf{8091x} \\
    \bottomrule
    \end{tabular}
    \caption{CPU time comparison on all test datasets. FastRP is over 4,000 times faster than the state-of-the-art algorithm DeepWalk.}
    \label{tab:runtime}
\end{table*}

\begin{table*}
\setlength\tabcolsep{2pt}
\caption{Top 5 nearest neighbors of four representative websites calculated from the node embeddings generated by \ouralgorithm{}, DeepWalk and RandNE respectively.
Rows are arranged from the highest cosine similarity to lowest cosine similarity.}
\label{tab:site-knns}
\small
\begin{tabular}{p{1.3cm}|lll | lll}
\toprule[0.1em]
Methods & \texttt{\ouralgorithm{}} & \texttt{DeepWalk} & \texttt{RandNE} & \texttt{\ouralgorithm{}} & \texttt{DeepWalk} & \texttt{RandNE} \\
\midrule 
Websites &
\multicolumn{3}{c|}{\texttt{nytimes.com}} &
\multicolumn{3}{c}{\texttt{delta.com}} \\
\midrule 
\multirow{5}{*}{Neighbors}
& {\small huffingtonpost.com} & {\small washingtonpost.com} & {\small huffingtonpost.com} & aa.com & aa.com & aa.com\\
& {\small washingtonpost.com} & {\small huffingtonpost.com} & {\small washingtonpost.com} & united.com & united.com & southwest.com\\
& cnn.com & cnn.com & forbes.com & usairways.com & usairways.com & united.com\\
& npr.org & cbsnews.com & cnn.com & alaskaair.com & southwest.com & expedia.com\\
& latimes.com & time.com & npr.org & jetblue.com & jetblue.com & priceline.com\\

\midrule[0.1em]

Methods & \texttt{\ouralgorithm{}} & \texttt{DeepWalk} & \texttt{RandNE} & \texttt{\ouralgorithm{}} & \texttt{DeepWalk} & \texttt{RandNE} \\
\midrule 
Websites &
\multicolumn{3}{c|}{\texttt{vldb.org}} &
\multicolumn{3}{c}{\texttt{arsenal.com}} \\
\midrule 
\multirow{5}{*}{Neighbors}
& sigmod.org & sigmod.org & comp.nus.edu.sg & chelseafc.com & chelseafc.com & liverpoolfc.com\\
& comp.nus.edu.sg & morganclaypool.com & cs.sfu.ca & mcfc.co.uk & {\small tottenhamhotspur.com} & manutd.com\\
& sigops.org & kdd.org & cs.rpi.edu & nufc.co.uk & manutd.com & chelseafc.com\\
& cidrdb.org & doi.acm.org & nlp.stanford.edu & avfc.co.uk & mcfc.co.uk & skysports.com\\
& cse.iitb.ac.in & informatic.uni-trier.de & theory.stanford.edu & {\small tottenhamhotspur.com} & thefa.com & {\small tottenhamhotspur.com}\\
\bottomrule[0.1em]
\end{tabular}
\end{table*}

\subsection{Qualitative Case Study: WWW-200K Network}
We first conduct a case study on the WWW-200K network to compare the embeddings produced by different network embedding algorithms qualitatively.
For this part, we take RandNE and DeepWalk as the baselines since the results produced by LINE and node2vec are very similar to those of DeepWalk on this network.

Examining the $K$-nearest neighbors (KNN) of a word is a common way to measure the quality of word embeddings~\cite{skipgram}.
In the same spirit, we examine the $K$-nearest neighbors of several representative websites in the node embedding space.
Cosine similarity is used as the similarity metric.
Table~\ref{tab:site-knns} lists the top 5 nearest neighbors of four representative websites: \texttt{nytimes.com}, \texttt{delta.com}, \texttt{vldb.org}, and \texttt{arsenal.com} based on the node embeddings produced by \ouralgorithm{}, RandNE and DeepWalk.

\texttt{nytimes.com} is the website of The New York Times, which is one of the most influential news websites in the US. We find that all three methods produce high-quality nearest neighbors for \texttt{nytimes.com}: \texttt{huffingtonpost.com}, \texttt{washingtonpost.com} and \texttt{cnn.com} are al\-so well-known American news sites.
It is also interesting to see that \ouralgorithm{} lists \texttt{latimes.com} (The Los Angeles Times) among the top 5 nearest neighbors of \texttt{nytimes.com}.

\texttt{delta.com} is the homepage of Delta Air Lines, which is a major American airline.
Again, we find that the most similar websites discovered by \ouralgorithm{} and DeepWalk are the official websites of other major American airlines: American Airlines (\texttt{aa.com} and \texttt{usairways.com}), United Airlines (\texttt{united.com}), Alaska Airlines (\texttt{alaskaair.com}), etc.
The nearest neighbors list provided by RandNE is worse, since it includes general purpose travel websites such as \texttt{expedia.com} and \texttt{priceline.com}.

\texttt{vldb.org} is the official website for VLDB Endowment, which steers the VLDB conference, a leading database research conference.
Both FastRP and DeepWalk list \texttt{sigmod.org} as the most similar website to \texttt{vldb.org}; this is a positive sign since SIGMOD is another top database research conference.
On the other hand, all three methods include several universities' CS department websites in the nearest neighbors list, such as \texttt{comp.nus.edu.sg} and \texttt{cs.sfu.ca}.
In particular, all top five websites provided by RandNE are CS department websites.
In our opinion, this is reasonable but less satisfactory than having other CS research conferences' websites in the list, such as \texttt{sigops.org} (The ACM Special Interest Group in Operating Systems
) and \texttt{cidrdb.org} (The Conference on Innovative Data Systems Research).

\texttt{arsenal.com} represents a football club that plays in the Premier League.
It can be seen that all three methods list the other football clubs in the Premier League as the nearest neighbors of \texttt{arsenal.com}, such as Chelsea (\texttt{chelseafc.com}) and Manchester City (\texttt{mcfc.co.uk}).
The only exception is RandNE, which also lists \texttt{skysports.com} (the dominant subscription TV sports brand in the United Kingdom) as one of the nearest neighbors.
\texttt{skysports.com} has higher popularity but is less relevant to \texttt{arsenal.com}.
In contrast, \ouralgorithm{} avoids this problem by properly downweights the influence of popular nodes.
Overall, we find that the quality of nearest neighbors produced by \ouralgorithm{} is comparable to DeepWalk and significantly better than that of RandNE.

The experiments above serve as a qualitative evaluation of \ouralgorithm{}.
In the following sections, we will conduct quantitative evaluations of \ouralgorithm{} on different downstream tasks across multiple datasets.

\begin{table}[!ht]
\centering
\begin{tabular}{clll}
\toprule
\textbf{Algorithm} & \multicolumn{3}{c}{\textbf{Dataset}}           		\\ & WWW-10K & BlogCatalog & Flickr \\
\midrule
\texttt{LINE} & 6.66 & 19.63 & 10.69 \\
\texttt{node2vec} & \textbf{27.42} & 21.44 & 11.89 \\
\texttt{DeepWalk} & 25.54 & 21.30 & 14.00 \\
\texttt{RandNE}   & 15.68 & 20.88 & 13.64\\
\texttt{\ouralgorithm{}} & 26.92 & \textbf{23.43} & \textbf{15.02} \\
\bottomrule
\end{tabular}
\caption{Macro $F_1$ scores of all methods on \texttt{WWW-10K}, \texttt{BlogCatalog}, and \texttt{Flickr} in percentage (Section \ref{sec:classfication}).
}
\label{tab:classification_summary}
\end{table}

\begin{figure*}[!ht]
    \centering
    \begin{subfigure}[b]{.3\linewidth}
        \includegraphics[width=\linewidth]{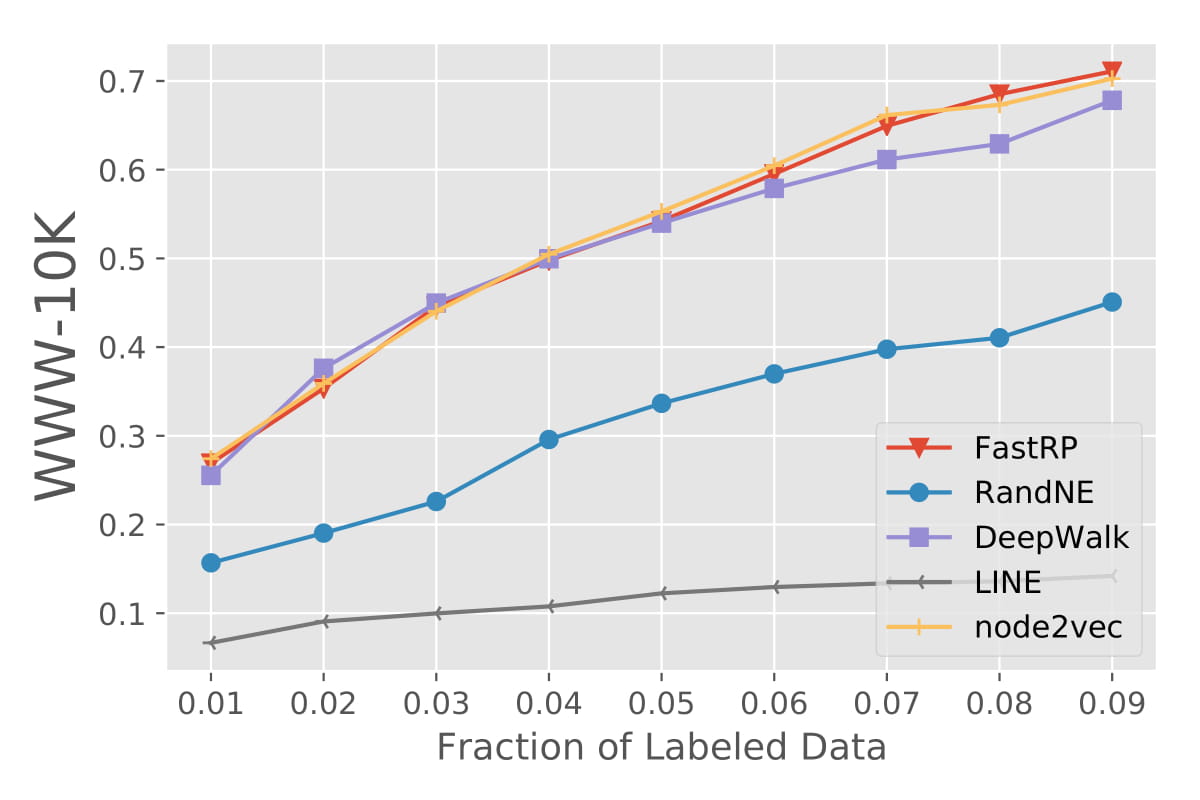}
	\end{subfigure}
    \begin{subfigure}[b]{.3\linewidth}
        \includegraphics[width=\linewidth]{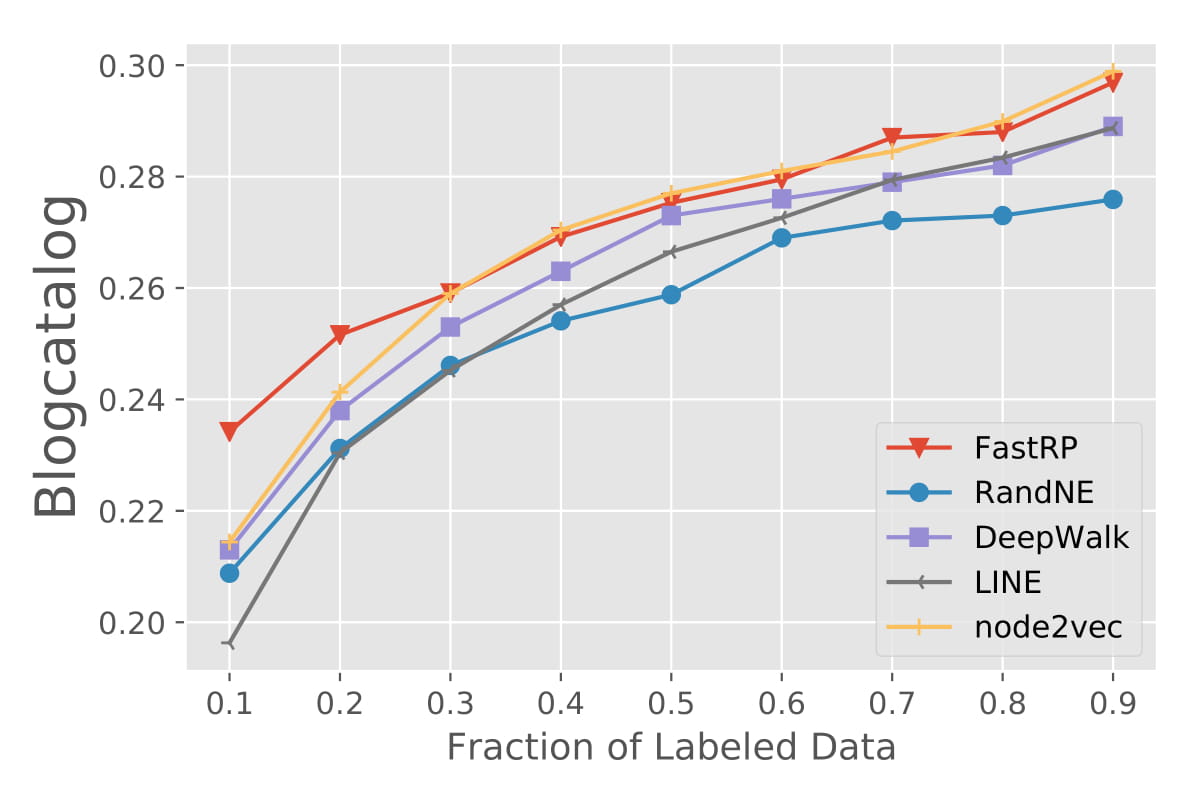}
    \end{subfigure}
    \begin{subfigure}[b]{.3\linewidth}
        \includegraphics[width=\linewidth]{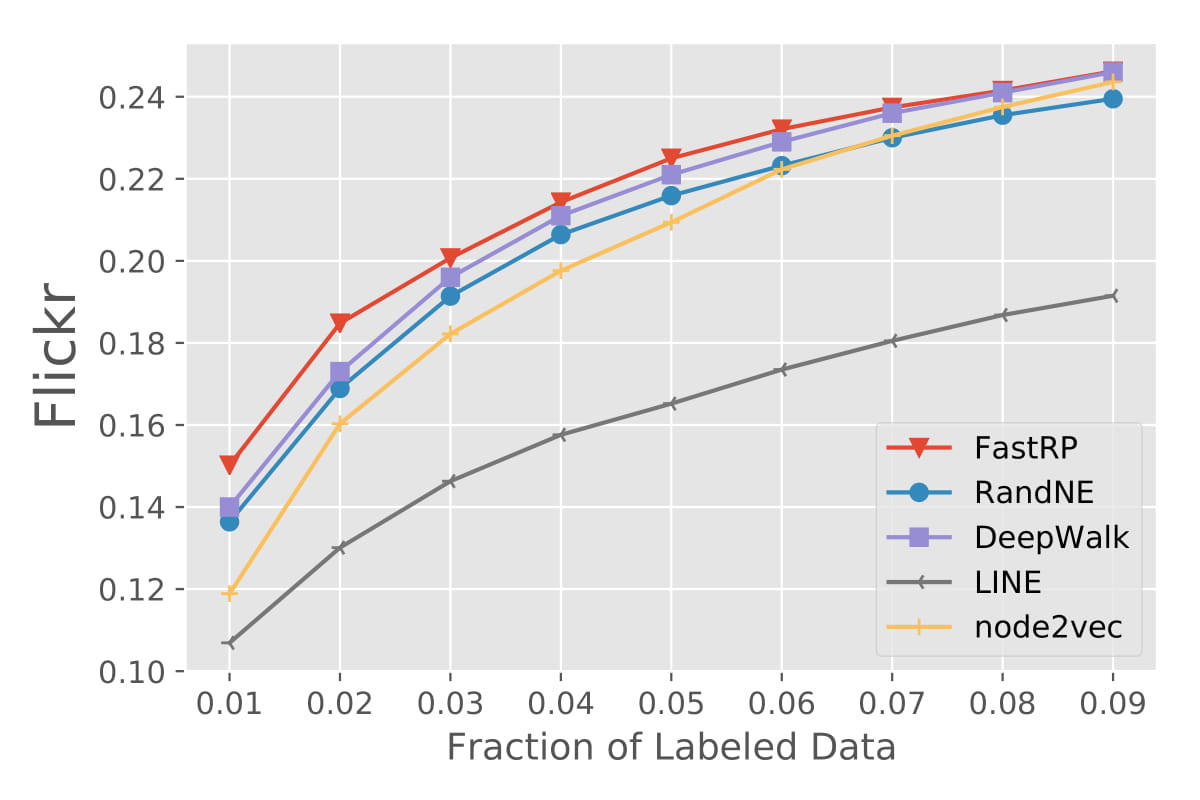}
    \end{subfigure}
    \caption{
    Detailed multi-label classification result on \texttt{WWW-10K}, \texttt{BlogCatalog}, and \texttt{Flickr} (Section \ref{sec:classfication}).
    }
    \label{fig:classification_details}

\end{figure*}

\subsection{Multi-label Node Classification}
\label{sec:classfication}

For the task of node classification, we evaluate our method using the same experimental setup in DeepWalk~\cite{deepwalk}.
Firstly, a portion of nodes along with their labels are randomly sampled from the graph as training data, and the goal is to predict the labels for the rest of the nodes.
Then, a one-vs-rest logistic regression model with L2 regularization is trained on the node embeddings for prediction.
We use the logistic regression model implemented by LibLinear~\cite{liblinear}.
To ensure the reliability of our experiment, the above process is repeated for 10 times, and the average Macro $F_1$ score is reported.
The other evaluation metrics such as Micro $F_1$ score and accuracy follow the same trend as Macro $F_1$ score, thus are not shown.

Table~\ref{tab:classification_summary} reports the Macro $F_1$ scores achieved on WWW-10K, Blogcatalog and Flickr with 1\%, 10\% and 1\% labeled nodes respectively.
\ouralgorithm{} achieves the state-of-the-art result on two out of three datasets and matches on the third.
On \texttt{Blogcatalog} and \texttt{Flickr}, \ouralgorithm{} achieves gains of 9.3\% and 7.3\% over the best performing baseline method respectively.
On \texttt{WWW-10K}, the absolute difference in $F_1$ score between \ouralgorithm{} and node2vec is only 0.5\%, but \ouralgorithm{} is over 10,000 times faster.

To have a detailed comparison between \ouralgorithm{} and the baseline methods, we vary the portion of labeled nodes for classification and present the macro $F_1$ scores in Figure~\ref{fig:classification_details}.
We can observe that \ouralgorithm{} consistently outperforms or matches the other neural baseline methods, while being at least three orders of magnitude faster.
It is also clear that \ouralgorithm{} always outperforms RandNE by a large margin, proving the effectiveness of our node similarity matrix design.

\subsection{Parameter Sensitivity}
\begin{figure*}[t]
\centering
\begin{subfigure}[b]{.3\linewidth}
	\centering
    \includegraphics[width=\linewidth]{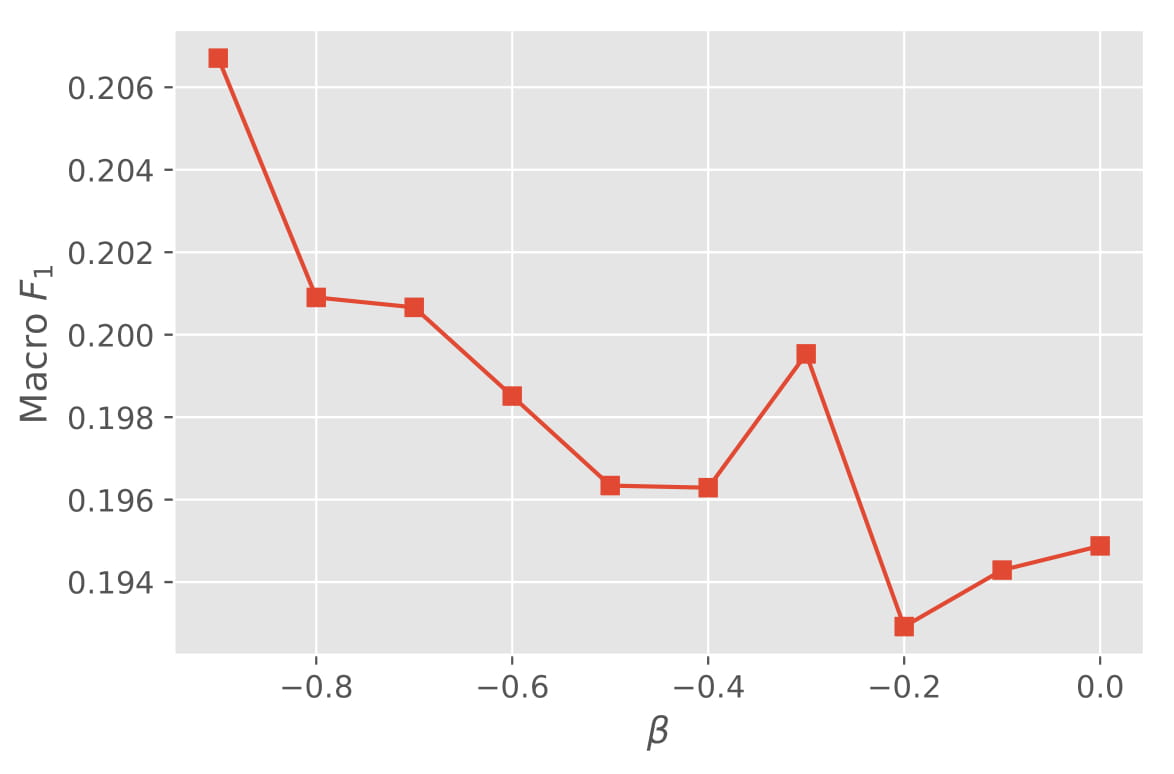}
    \caption{Normalization strength $\beta$.}
    \label{fig:param-beta}
\end{subfigure}
\begin{subfigure}[b]{.3\linewidth}
	\centering
    \includegraphics[width=\linewidth]{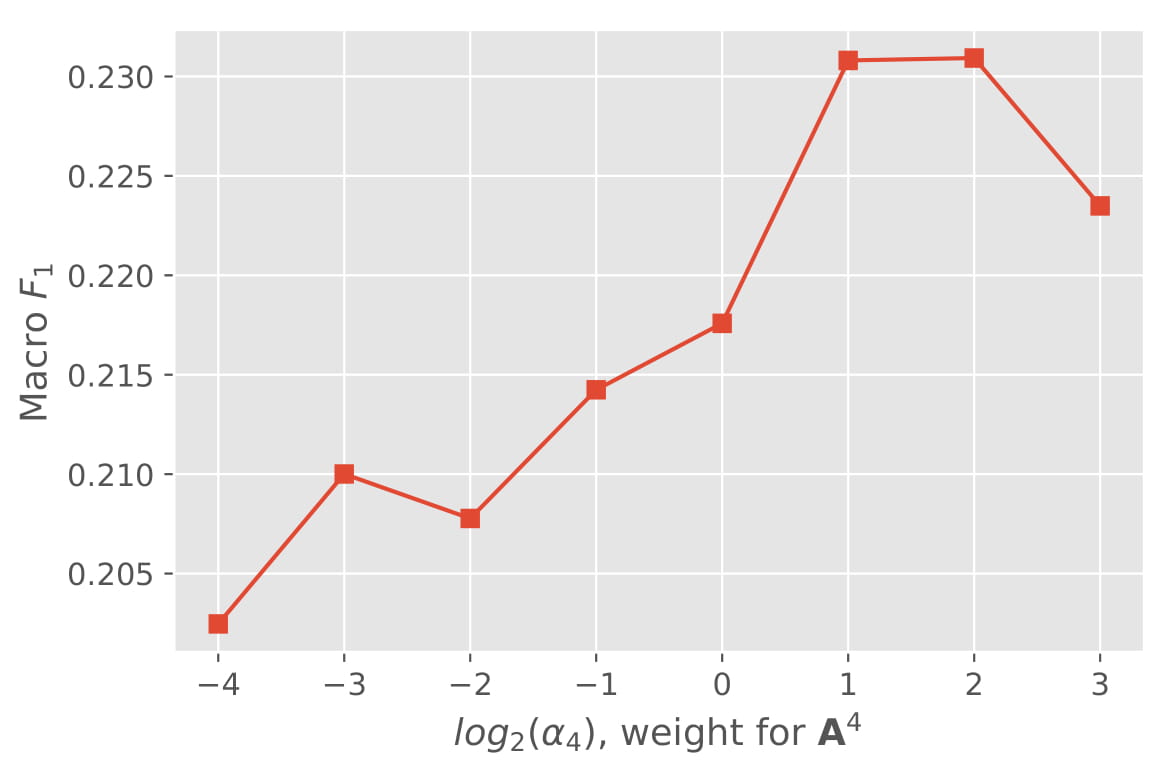}
    \caption{Weight $\alpha_4$.}
    \label{fig:param-weight}
\end{subfigure}
\begin{subfigure}[b]{.3\linewidth}
	\centering
    \includegraphics[width=\linewidth]{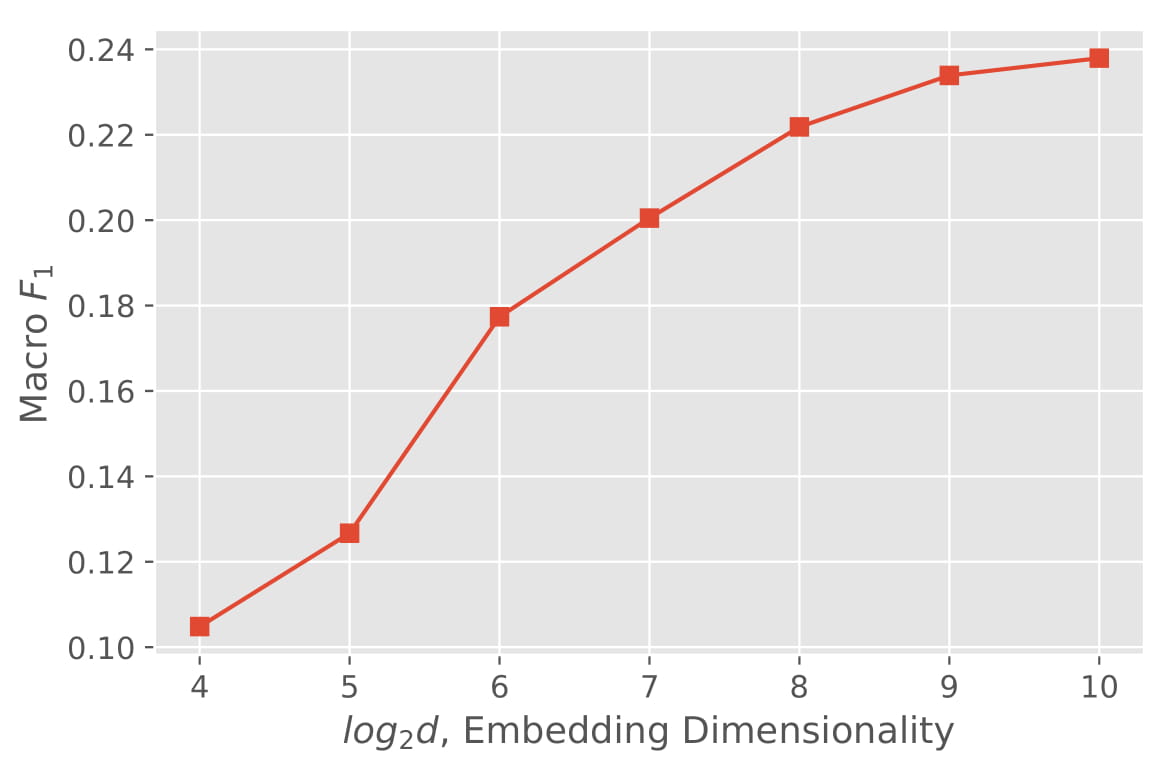}
    \caption{Embedding dimensionality $d$.}
    \label{fig:param-dim}
\end{subfigure}
\caption{Parameter sensitivity study on \texttt{Blogcatalog}.}
\label{fig:param-sensitivity}
\end{figure*}

To examine how do the hyperparameters affect the quality of learned representations,
we conduct a parameter sensitivity study on \texttt{BlogCatalog}.
The parameters we investigate are the normalization strength $\beta$, $\mathbf{A}^4$'s weight $\alpha_4$, and the embedding dimensionality $d$.
We report the Macro $F_1$ score achieved with 10\% labeled data in Figure~\ref{fig:param-sensitivity}.

\noindent \textbf{Normalization strength.} Figure~\ref{fig:param-beta} shows the effectiveness of our normalization scheme. At $\beta = -0.9$, \ouralgorithm{} achieves the highest $F_1$ score of over 20.6\%.
By setting $\beta$ to $0.0$, no normalization is applied to the node similarity matrix and the $F_1$ score drops to 19.5\%. We set $d$ to 128 for this experiment.

\noindent \textbf{Weight $\alpha_4$}. Figure~\ref{fig:param-weight} shows that the weight $\alpha_4$ also plays an important role. The best $F_1$ score is achieved when $\alpha_4$ is set to $2$ or $4$ on this dataset.

\noindent \textbf{Embedding Dimensionality.} Figure~\ref{fig:param-dim} shows that increasing the embedding dimensionality in general yields better node embeddings.
On the other hand, we notice that \ouralgorithm{} already achieves better performance than all baseline method with embedding dimensionality of 256.

\subsection{Scalability}
\begin{figure}[t]
\centering
\begin{subfigure}[b]{.48\linewidth}
	\centering
    \includegraphics[width=\linewidth]{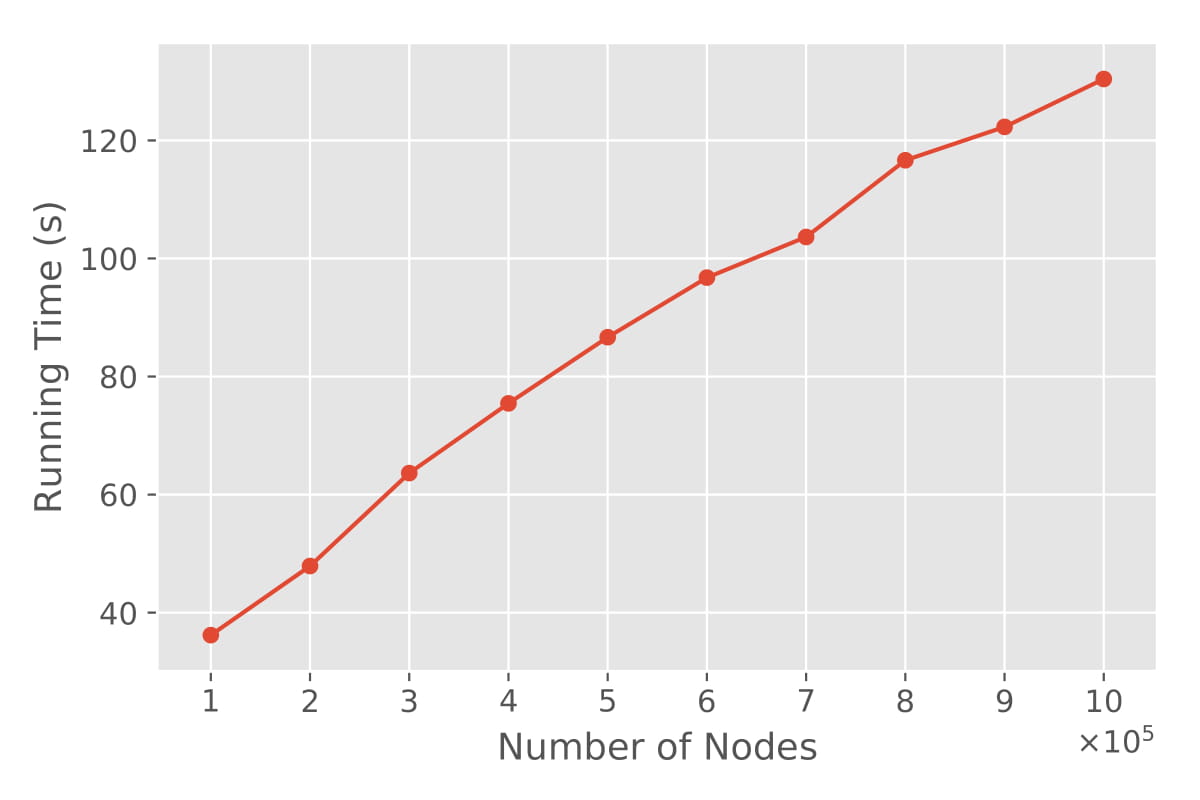}
    \caption{Change $n$.}
    \label{fig:runtime-n}
\end{subfigure}
\begin{subfigure}[b]{.48\linewidth}
	\centering
    \includegraphics[width=\linewidth]{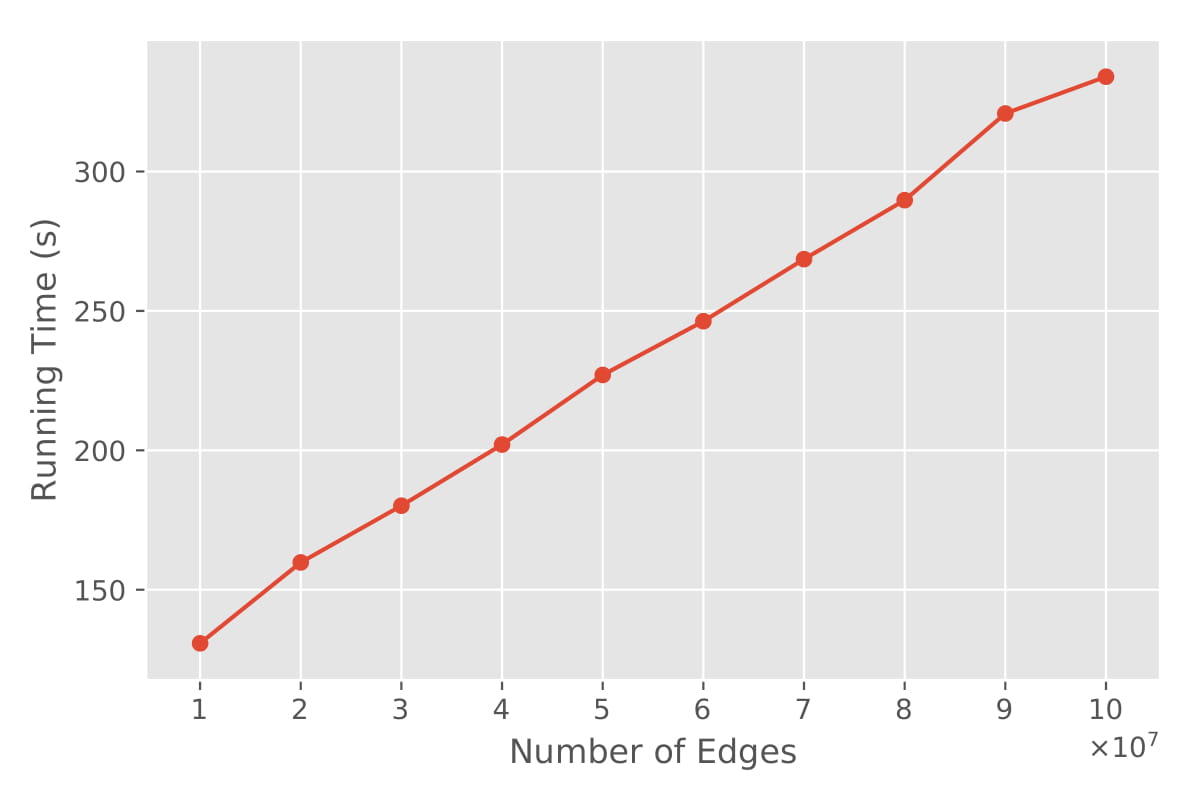}
    \caption{Change $m$.}
    \label{fig:runtime-m}
\end{subfigure}
\caption{Scalability study on Erdos-Renyi graphs.}
\label{fig:runtime}
\end{figure}

In Section~\ref{sec:time-complexity}, we show that the time complexity of \ouralgorithm{} is linear to the number of nodes $n$ and the number of edges $m$.
Here, we empirically verifies this by learning embeddings on random graphs generated by the Erdos-Renyi model.
In Figure~\ref{fig:runtime-n}, we fix $m$ to $10^7$ and vary $n$ from $10^5$ to $10^6$.
In Figure~\ref{fig:runtime-m}, we fix $n$ to $10^6$ and vary $m$ from $10^7$ to $10^8$.
For both figures, we report the CPU time for \ouralgorithm{} to embed the graph.
It can be seen that the empirical running time of \ouralgorithm{} also scales linearly with $n$ and $m$.

\section{Related Work}
\noindent \textbf{Network embeddings.}
Most of the early methods view network embeddings as a dimension reduction problem,
where the goal is to preserve the local or global distances between data points in a low-dimensional manifold~\cite{lle,isomap,le}. With time complexity at least quadratic in the number of data points (or nodes), these methods do not scale to graphs with hundreds of thousands of nodes.

Inspired by the success of scalable neural methods for learning word embeddings,
in particular the Skip-gram model~\cite{skipgram}, neural methods are proposed for network embeddings~\cite{deepwalk,line,node2vec}.
These methods typically sample node pairs that are close to each other
and then train a Skip-gram model on the pairs to obtain node embeddings.
The difference mostly lies in the strategy for node pairs sampling.
DeepWalk~\cite{deepwalk} samples node pairs that are at most $k$ hops away via random walking on the graph.
Node2vec~\cite{node2vec} introduces a biased random walk strategy using a mixture of DFS and BFS.
LINE~\cite{line} considers the node pairs that are 1-hop or 2-hops away from each other.
These methods not only produce high-quality node embeddings but also scale to networks with millions of nodes. 

In Levy and Goldberg's seminal work~\cite{word-emb-as-mf} on interpreting Skip-gram with negative sampling (SGNS), they prove that SGNS implicitly factorizes a shifted pointwise mutual information (PMI) matrix of word co-occurrences.
Using a similar methodology, it is shown that methods like DeepWalk~\cite{deepwalk}, LINE~\cite{deepwalk}, PTE~\cite{pte} and node2vec~\cite{node2vec} all implicitly approximate and
factorize a node similarity matrix, which is usually some transformation of the $k$-step transition matrices $\mathbf{A}^k$~\cite{tadw,netmf} .
Following these analyses, matrix factorization-based methods are also proposed for network embeddings~\cite{tadw,netmf,grarep}.
A representative method is GraRep~\cite{grarep}, which can be seen as the matrix factorization version of DeepWalk: it uses SVD to factorize the shifted PMI matrix of $k$-step transition matrices. However, GraRep is not scalable due to the high time complexity of both raising the transition matrix $A$ to higher powers and taking the element-wise logarithm of $\mathbf{A}^k$, which is a dense $n \times n$ matrix.

A few recent work thus propose to speed up the construction of such a node similarity matrix~\cite{arope,netmf,netsmf}, which are inspired by the spectral graph theory.
The basic idea is that if the top-$h$ eigendecomposition of $\mathbf{A}$ is given by $\mathbf{A} = \mathbf{U}_{h} \mathbf{\Lambda}_{h} \mathbf{U}_h^{\top}$, then $\mathbf{A}^k$ can be approximated with $\mathbf{U}_{h} \mathbf{\Lambda}_{h}^k \mathbf{U}_h^{\top}$~\cite{netmf,arope}.
The major drawback of these methods is that they need to get rid of the element-wise normalization (such as taking logarithm) on $A^k$ to achieve better scalability; this harms the quality of embeddings~\cite{gemd}. 
A recent method~\cite{netsmf} proposes to sparsify $\mathbf{A}^k$ for better scalability.
However, the similarity matrix is still dense even after the sparsification: for a graph with $n=10^6$ nodes and $m=10^7$ edges, the number of entries in the sparsified matrix can be as high as $1.4 \times 10^{11}$~\cite{netsmf}.

Perhaps the most relevant work is RandNE~\cite{randne}, which considers a Gaussian random projection approach for network embeddings.
There are three key differences between our work and RandNE:
\begin{enumerate}[wide, labelwidth=!, labelindent=0pt]
\item We are the first to identify the two key factors for constructing the node similarity matrix: high-order proximity preservation and element normalization.
Specifically, the importance of normalization is overlooked in many previous studies, including RandNE~\cite{randne,netmf,arope}.
\item Based on theoretical analysis, we derive a normalization algorithm that properly downweights the influence of high-degree nodes in the node similarity matrix.
An additional advantage of our normalization approach is that it can be formalized as a simple matrix multiplication operation,
which enables fast iterative computation when combined with random projection.
\item We explore the usage of very sparse random projection for network embeddings, which is more efficient than traditional Gaussian random projection.
As shown in the experiments, \ouralgorithm{} achieves substantially better performance on challenging downstream tasks while being faster.
\end{enumerate}

\noindent \textbf{Graph-based Recommendation Systems.}
Our work is also related to graph-based recommendation systems, which consider a special kind of graph: the bipartite graph between users and items.
Typically, the goal is to generate the top-K items that a user will be most interested in.

Several early work in this field emphasis on the importance of high-order, transitive relationships between users and items~\cite{p3-1,p3-2,p3-alpha,rp3} for top-K recommendation.
Fouss et al.~\cite{p3-1,p3-2} present $P^3$, which directly uses the entries in the third power of the transition matrix to rank items.
Although achieving competitive recommendation performance, it is observed that the ranking of items in $P^3$ is strongly influenced by the popularity of items~\cite{p3-alpha,rp3}: popular items tend to dominate the recommendation list for most users.
To this end, $P^3_{\alpha}$~\cite{p3-alpha} and $RP^3_{\beta}$~\cite{rp3} are proposed as re-weighted versions of $P^3$.
Their idea of downweighting popularity items is similar to the normalization strategy in this paper.
However, these methods are proposed as heuristics specifically for bipartite graphs and the task of top-K recommendation, which do not generalize to other scenarios.
Moreover, the power of the transition matrix is either computed exactly~\cite{p3-alpha} or approximated by sampling a significant number of random walks~\cite{rp3},
both of which are not scalable.

\section{Conclusion}
We present \ouralgorithm{}, a scalable algorithm for obtaining distributed representations of nodes in a graph.
\ouralgorithm{} first constructs a node similarity matrix that captures high-order proximity between nodes and then normalizes the matrix entries based on the convergence properties of $\mathbf{A}^k$.
Very sparse random projection is applied to this similarity matrix to obtain node embeddings.
Experimental results show that \ouralgorithm{} achieves three orders of magnitudes speedup over state-of-the-art method DeepWalk while producing embeddings of comparable or even better quality.

\section{Acknowledgements}
This work was partially supported by NSF grants IIS-1546113 and IIS-1927227.

%
\bibliographystyle{ACM-Reference-Format}
\bibliography{main}

\end{document}